\documentclass[aps,prl,twocolumn,showpacs]{revtex4}

\usepackage{epsfig}
\usepackage{graphicx}
\usepackage{amsmath}
\usepackage{bm}

\newcommand{\ltsim}{\protect\raisebox{-0.5ex}{$\:\stackrel{\textstyle <}{\sim}\:$}}
\newcommand{\gtsim}{\protect\raisebox{-0.5ex}{$\:\stackrel{\textstyle >}{\sim}\:$}}

\begin{document}
\setlength{\columnwidth}{0.7\textwidth}

\title{Anomalous temperature dependence of the single-particle
spectrum in the organic conductor TTF-TCNQ}

\author{N. Bulut$^{a,b}$, H. Matsueda$^{a,c}$, T. Tohyama$^a$, and S. Maekawa$^{a,b}$}

\address{$^a$Institute for Materials Research, Tohoku University, Sendai
980-8577, Japan\\
$^b$CREST, Japan Science and Technology Agency (JST), Kawaguchi,
Saitama 332-0012, Japan\\
$^c$Department of Physics, Tohoku University, Sendai 980-8578,
Japan}

\date{March 5, 2006}

\begin{abstract}
The angle-resolved photoemission spectrum of the organic conductor
TTF-TCNQ exhibits an unusual transfer of spectral weight over a
wide energy range for temperatures $60K<T<260K$. In order to
investigate the origin of this finding, here we report numerical
results on the single-particle spectral weight $A(k,\omega)$ for
the one-dimensional (1D) Hubbard model and, in addition, for the
1D extended Hubbard and the 1D Hubbard-Holstein models.
Comparisons with the photoemission data suggest that the 1D
Hubbard model is not sufficient for explaining the unusual $T$
dependence, and the long-range part of the Coulomb repulsion also
needs to be included.
\end{abstract}

\pacs{71.10.Fd, 79.60.Fr, 71.20.Rv, 72.15.Nj}

\maketitle

The low-dimensional interacting systems receive attention because
of their unusual electronic properties \cite{Voit}. In this
respect, the high-resolution Angle-Resolved Photoemission
Spectroscopy (ARPES) measurements on the quasi-one-dimensional
organic conductor tetrathiofulvalene-tetracyanoquinodimethan
(TTF-TCNQ) have provided evidence for non-Fermi liquid behavior in
this compound \cite{Zwick,Claessen,Sing,Ito}. In particular, the
ARPES experiments have found that the single-particle spectral
weight at the Fermi wavevector $k_F$ is transferred over an energy
range of $\approx 1.3eV$ of the Fermi level in the TCNQ-derived
band, as the temperature $T$ decreases from $260K$ to $60K$
\cite{Claessen,Sing}. In a Fermi liquid the spectral-weight
transfer would have occurred within $\sim k_BT$ of the Fermi
level. Here, we investigate the origin of this unusual ARPES data
and its meaning for the electronic structure of TTF-TCNQ by using
the Dynamical Density Matrix Renormalization Group (DDMRG),
Quantum Monte Carlo (QMC) and the exact diagonalization methods.

There are various possibilities as to what might be the origin of
the anomalous $T$ dependence of the single-particle spectral
weight at the Fermi level, $A(k_F,\omega)$, in TTF-TCNQ: (i) It
has been suggested that the $T$-dependence of the ARPES data can
be explained within the one-dimensional (1D) Hubbard model
\cite{Claessen,Sing}. In this case, the anomalous $T$-dependence
of $A(k_F,\omega)$ over the conduction bandwidth has been
attributed to the strong-correlation effects. Indeed, by using the
Bethe-Ansatz solution, the photoemission spectrum has been fitted
excellently to the dispersion of the spinon and holon bands of the
1D Hubbard model with the parameters $t=0.4eV$ for the hopping
matrix element and $U=2eV$ for the Coulomb repulsion. The recent
observation of the $3k_F$ structures in $A(k,\omega)$ by the ARPES
\cite{Ito} also supports this picture. (ii) An alternative point
of view is that an extended Hubbard model with long-range Coulomb
repulsion is necessary, particularly because the screening of the
long-range Coulomb repulsion is expected to be weaker for the
surface layer of TTF-TCNQ. (iii) Another possibility is that the
electron-phonon interaction, in addition to the strong Coulomb
repulsion, plays a role in producing the unusual $T$ dependence.
In this paper, our goal is to differentiate among these
possibilities. For this purpose, we present DDMRG and
finite-temperature QMC results on $A(k_F,\omega)$ of the 1D
Hubbard model. In addition, we present exact-diagonalization
results for the 1D extended Hubbard model which includes a
near-neighbor repulsion $V$ and DDMRG results for the 1D
Hubbard-Holstein model.

In the following, we show that, above a characteristic temperature
determined by the effective magnetic exchange $J_{eff}$,
spectral-weight transfer takes place over a wide energy range in
the 1D Hubbard model. This is similar to the $T$ dependence
observed in the ARPES experiments. However, below this
temperature, the weight transfer is negligible. We find that the
Hubbard parameters $t=0.4eV$ and $U=2eV$ give too large a value
for $J_{eff}$, and with these parameters it is not possible to
explain the ARPES $T$ dependence. On the other hand, the $T=0$
exact-diagonalization results on the 1D extended Hubbard model
show that the nearest-neighbor Coulomb repulsion increases the
bandwidth of the spinon and the holon excitations. This can lead
to a smaller value for $t$ for fitting the ARPES dispersions, and
a reduced value for $J_{eff}$, hence, giving better agreement with
the ARPES data. In addition, the DDMRG results on the
Hubbard-Holstein model show that, at $T=0$, the electron-phonon
interaction influences $A(k_F,\omega)$ only at $|\omega|\ll 1.3eV$
for physical values of the phonon frequency $\omega_0$. However,
it remains to be seen how the electron-phonon interaction
influences $A(k,\omega)$ at finite $T$. The main finding of this
paper is that the 1D Hubbard model is not sufficient for
explaining simultaneously the $T$ dependence and the dispersion of
the photoemission spectrum of TTF-TCNQ. We suggest that it is
necessary to include at least the long-range part of the Coulomb
repulsion.

The Hubbard hamiltonian $H_0$ is defined by
\begin{eqnarray}
H_0&=&-t\sum_{i,\sigma}( c^{\dagger}_{i,\sigma}c_{i+1,\sigma} +
{\rm h.c.} )+U\sum_{i}n_{i,\uparrow}n_{i,\downarrow}
\end{eqnarray}
where $c_{i,\sigma}$ ($c_{i,\sigma}^{\dagger}$) annihilates
(creates) an electron with spin $\sigma$ at lattice site $i$,
$n_{i}=n_{i,\uparrow}+n_{i,\downarrow}$,
$n_{i,\sigma}=c_{i,\sigma}^{\dagger}c_{i,\sigma}$, $t$ is the
hopping integral, and $U$ is the on-site Coulomb repulsion. We
will consider the case of electron-filling $\langle
n\rangle=0.60$, since the filling of the TCNQ band is 0.59. We
note that $A(k,\omega)$ of the 1D Hubbard model was studied with
the QMC \cite{Preuss,Zacher,Matsueda,Abendschein} and the DDMRG
\cite{Bethien,Matsueda} as well as with the Bethe ansatz
\cite{Penc}.

Within the DDMRG method, the single-particle spectral weight is
obtained at $T=0$ from
\begin{eqnarray}
A(k,\omega)=-\frac{1}{\pi}{\rm
Im}\left<0\left|c^{\dagger}_{k,\uparrow}\frac{1}{E_{0}-\omega-H+i\gamma}
c_{k,\uparrow}\right| 0\right>
\end{eqnarray}
where $c_{k,\uparrow}$ annihilates an electron with wavevector $k$
and spin $\uparrow$, $\left|0\right>$ and $E_{0}$ are the ground
state and the eigenenergy, respectively, and $\gamma$ is a small
positive number. The DDMRG results were obtained with the open
boundary conditions. At finite temperatures, we obtain
$A(k,\omega)$ by using the determinantal QMC technique
\cite{White}. This method yields the single-particle Green's
function along the Matsubara time axis, from which $A(k,\omega)$
is obtained by the maximum-entropy analytic continuation
\cite{Linden}. We have checked the convergence of the
maximum-entropy results as the statistics of the QMC data
improved. In the following, $A(k,\omega)$ will be plotted in units
of $t^{-1}$. In order to determine the characteristic temperature
of the magnetic correlations, we present QMC data on the uniform
magnetic susceptibility $\chi(q\rightarrow 0)$ where
\begin{equation}
\chi(q) = \int_0^{\beta} d\tau \sum_{\ell} e^{-iq\ell} \langle
m^z_{i+\ell}(\tau) m^z_i(0) \rangle
\end{equation}
with $m^z_i = n_{i,\uparrow}-n_{i,\downarrow}$.

We also present exact-diagonalization results on $A(k,\omega)$ for
the 1D extended Hubbard model,
\begin{eqnarray}
H_{ext}=H_0 +V\sum_{i}n_{i}n_{i+1}
\end{eqnarray}
where $V$ is the nearest-neighbor Coulomb repulsion. The
$A(k,\omega)$ of the 1D extended Hubbard model was previously
studied with the exact diagonalization technique \cite{Yunoki}. In
addition, we present DDMRG results on $A(k_F,\omega)$ of the
Hubbard-Holstein model defined by
\begin{eqnarray}
H_{HH}=H_0
+\omega_{0}\sum_{i}b^{\dagger}_{i}b_{i}+g\sum_{i}(b^{\dagger}_{i}
+ b_{i})n_{i} \label{HH}
\end{eqnarray}
where $b^{\dagger}_{i}$ ($b_{i}$) is the creation (annihilation)
operator for a dispersionless phonon at site $i$, $\omega_{0}$ is
the phonon frequency, and $g$ is the electron-phonon coupling
constant.

\begin{figure}[t]
\centering
\includegraphics[width=7cm,bbllx=100,bblly=257,bburx=525,bbury=686,clip]{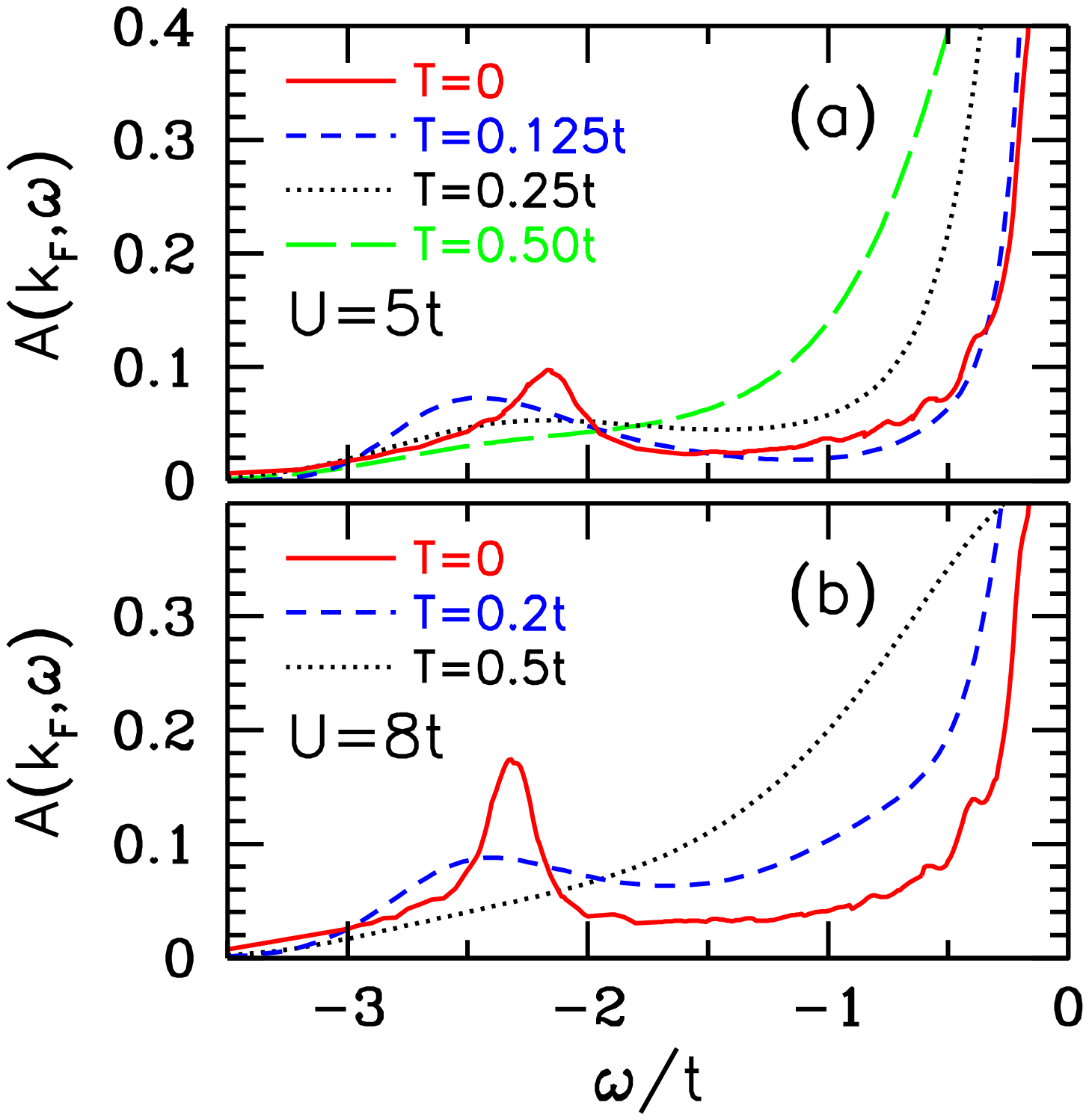}\\
\includegraphics[width=7cm,bbllx=100,bblly=261,bburx=525,bbury=515,clip]{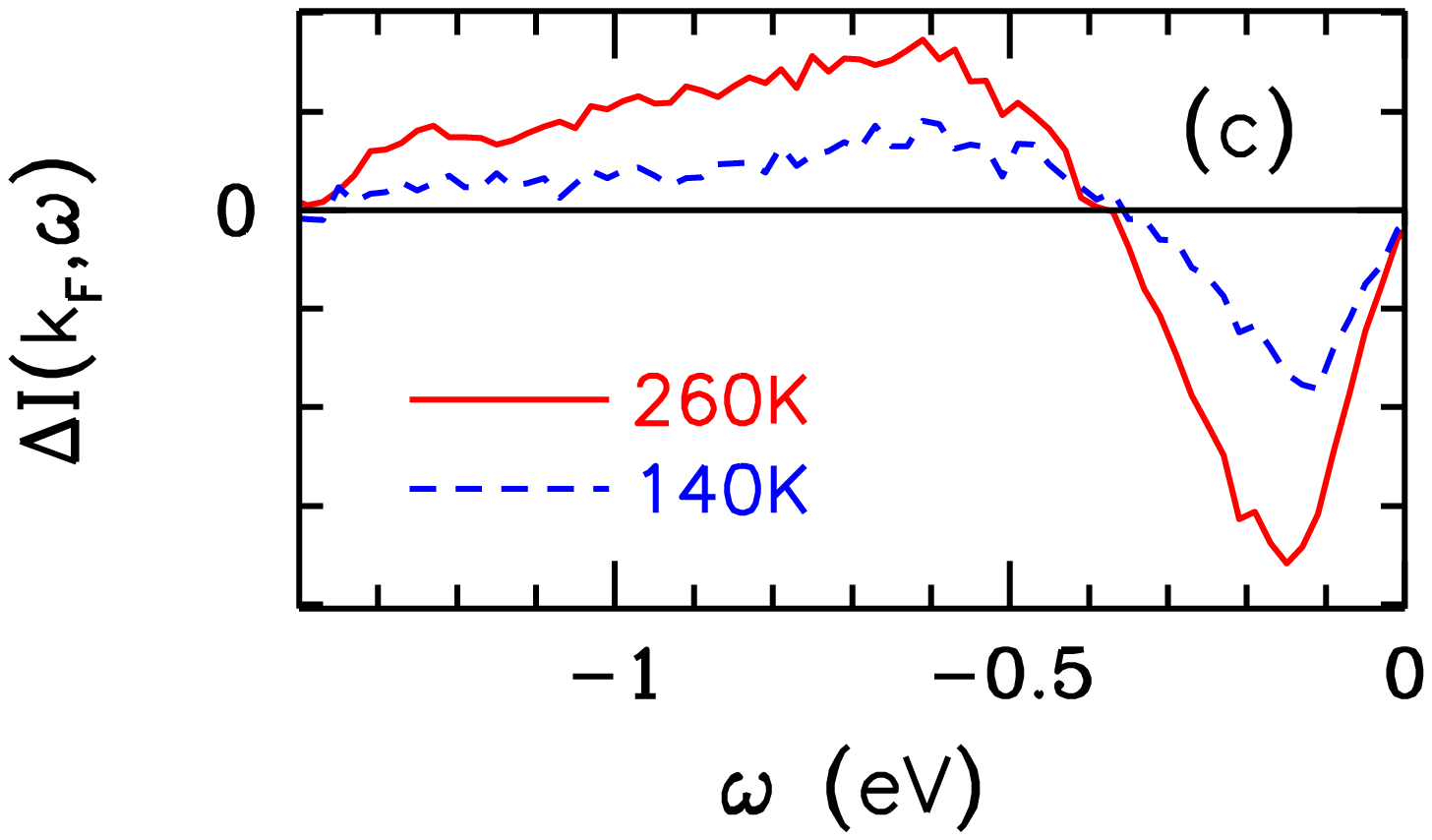}
\caption{(color online) Temperature dependence of the
single-particle spectral weight at the Fermi wavevector,
$A(k_F,\omega)$, for the 1D Hubbard model with (a) $U=5t$ and (b)
$U=8t$. In (c), the change in the ARPES intensity with respect to
$T=60K$, $\Delta I(k_F,\omega)$, is plotted (in arbitrary units)
for TTF-TCNQ (reproduced from Refs. \cite{Claessen,Sing} with
permission). } \label{fig1}
\end{figure}
We first discuss DDMRG and QMC results on the 1D Hubbard model.
Figures 1(a) and (b) show the $T$ dependence of $A(k_F,\omega)$
for $U=5t$ and $8t$. The DDMRG results were obtained for a 60-site
chain with 36 electrons, while the QMC results are for a 32-site
ring with $\langle n\rangle=0.60$. In the DDMRG calculations we
have used a finite energy broadening $\gamma=0.05t$. For $U=5t$,
we observe that the spectral-weight transfer occurs over an energy
range $\approx 2t$ of the Fermi level, as $T$ decreases from
$0.5t$ down to $0.125t$. However, the weight transfer is
negligible between $T=0.125t$ and $T=0$. For $t \approx 400meV$
\cite{Claessen,Sing}, $T=0.125t$ corresponds to $\approx 600K$.
Hence, in this case, the amount of spectral-weight transfer
between $T=600K$ and $T=0K$ is negligible, which disagrees with
the ARPES results. In order to study the dependence on $U/t$, in
Fig. 1(b) we show results for $U=8t$. Comparison of Figs. 1(a) and
(b) shows that the transfer of weight at low $T$ is enhanced for
$U=8t$ with respect to $U=5t$. In Fig. 1(c) we show the change in
the ARPES intensity at $k_F$, $\Delta I(k_F,\omega)$, for TTF-TCNQ
\cite{Claessen,Sing}. These results represent the photoemission
intensity arising mainly from the TCNQ-derived band. Here, the
solid curve represents the difference in $I(k_F,\omega)$ between
$260K$ and $60K$. The transfer of intensity is also observable as
$T$ is lowered from $140K$ to $60K$ (dashed curve). It is
considered that, at low energies $|\omega|\lesssim 0.4eV$, the
interchain hopping becomes important \cite{Claessen,Sing}. Hence,
here we will discuss the energy range $-1.3 eV \lesssim \omega
\lesssim -0.4eV$. In Figs. 1(a)-(c), we observe important
differences between the ARPES data and the numerical results. With
$U=5t$, it is not possible to explain the low temperature scale of
the weight transfer. In addition, we observe that it is not
possible to explain the energy range. In the ARPES data, the
intensity at $-1.3eV \lesssim \omega \lesssim -0.4eV$ decreases
with decreasing $T$. In the 1D Hubbard model and for $t=0.4eV$,
the decrease of the spectral weight occurs for $-0.8eV \lesssim
\omega \lesssim 0$, while the holon peak develops at
$\omega\approx -0.85eV$. Hence, it is not possible to explain the
$T$ and $\omega$ dependence of the ARPES data using the 1D Hubbard
model.

\begin{figure}[t]
\centering
\includegraphics[width=6.3cm,bbllx=100,bblly=204,bburx=525,bbury=512,clip]{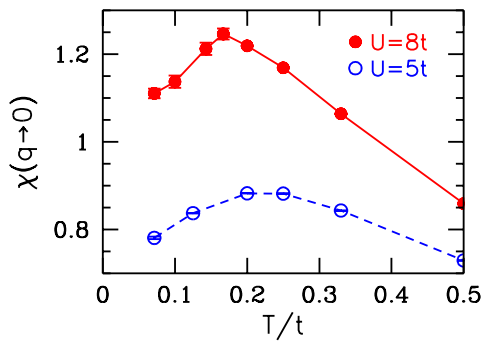}
\caption{(color online) Temperature dependence of the uniform
magnetic susceptibility $\chi(q\rightarrow 0)$, plotted in units
of $t^{-1}$, for the 1D Hubbard model. We use these data to
determine the effective magnetic exchange. } \label{fig2}
\end{figure}
In order to determine the effective magnetic exchange for the 1D
Hubbard model, in Fig. 2 we show the $T$ dependence of the uniform
magnetic susceptibility $\chi(q\rightarrow 0)$. We obtain a value
for $J_{eff}$ by making comparisons with the 1D Heisenberg model.
In the 1D Heisenberg model, as $T$ is lowered, $\chi(q\rightarrow
0)$ starts to decrease with the development of the $2k_F$ magnetic
correlations, and the maximum of $\chi(q\rightarrow 0)$ occurs at
$T_m\approx 0.64 J_{eff}$ \cite{BF}. Obtaining $T_m$ from Fig. 2,
we find $J_{eff}\approx 0.35t$ and $\approx 0.26t$ for $U=5t$ and
$8t$, respectively. Hence, we have $J_{eff}\approx 1600K$ for
$t=0.4eV$ and $U=2eV$. The results displayed in Fig. 1(a) for
$U=5t$ show that the transfer of weight over the wide energy range
occurs for $T \gtsim J_{eff}/3$, implying that this process
depends on the development of the short-range magnetic
correlations. Apparently, the long-range magnetic correlations are
not playing a role in this case. At this point, we also note that
$\chi(q\rightarrow 0)$ of bulk TTF-TCNQ is strongly suppressed by
the fluctuations of the Charge-Density-Wave (CDW) gap as $T$
decreases from the room temperature down to the CDW transition
temperature $T_{CDW}=53K$ \cite{Scott,Torrance}.

\begin{figure}[t]
\centering
\includegraphics[width=6.3cm,bbllx=100,bblly=204,bburx=525,bbury=512,clip]{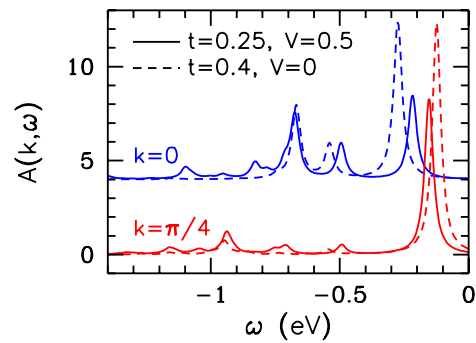}
\caption{(color online) Exact-diagonalization results on the
single-particle spectral weight $A(k,\omega)$ at $k=0$ and
$k=\pi/4$ for the 1D extended Hubbard model which has onsite and
nearest-neighbor Coulomb repulsions $U$ and $V$, respectively.
These calculations were performed for $U=2eV$ on a 16-site ring
with 10 electrons corresponding to $\langle n\rangle=0.625$. Here,
spectra obtained for $t=0.25eV$ and $V=0.5eV$ are compared with
that for $t=0.4eV$ and $V=0$.} \label{fig3}
\end{figure}
The density-functional-theory calculations deduce that the hopping
parameter $t=0.175eV$ for bulk TTF-TCNQ, while the analysis of the
ARPES data yields the Hubbard parameters $t=0.4eV$ and $U=5t$
\cite{Sing}. This enhancement of $t$ for the surface layer has
been attributed to a possible tilting of the TCNQ and TTF
molecules at the surface. In this paper, we have seen that the
parameters $t=0.4eV$ and $U=2eV$ give $J_{eff}\approx 1600K$,
which is too high to explain the $T$ dependence of $A(k,\omega)$.
At this point, we suggest that the long-range part of the Coulomb
repulsion might play an important role. For demonstration, we
present exact-diagonalization results on $A(k,\omega)$ for the 1D
extended Hubbard model with $U=2eV$. Figure 3 compares
$A(k,\omega)$ obtained for $t=0.25eV$ and $V=0.5eV$ with that for
$t=0.4eV$ and $V=0$ at wavevectors $k=0$ and $k=\pi/4$. For the
16-site lattice, $k=\pi/4$ is the closest wavevector to $k_F$. We
observe that, at $k=0$ and for $t=0.4eV$ and $V=0$, the holon and
spinon branches are located at $\approx -0.68eV$ and $\approx
-0.27eV$, respectively. For parameters $t=0.25eV$ and $V=0.5eV$,
these structures are located at similar energies. Hence, it is
possible to reproduce the locations of the spinon and holon
branches by using a reduced value for $t$ within the 1D extended
Hubbard model. This behavior is also observed at $k=\pi/4$. Figure
3 also shows that $V$ induces incoherent spectral weight at higher
$|\omega|$. For the 1D extended Hubbard model, we expect
$J_{eff}\propto 4t^2/(U-V)$. These results suggest that taking
into account the long-range part of the Coulomb repulsion in
fitting the ARPES dispersion can lead to a reduced $t$ and, hence,
might reduce $J_{eff}$ and the characteristic temperature for the
single-particle weight transfer. However, it is necessary to
calculate $A(k,\omega)$ for the 1D extended Hubbard model at
finite $T$.

\begin{figure}[t]
\centering
\includegraphics[width=6.3cm,bbllx=100,bblly=204,bburx=525,bbury=512,clip]{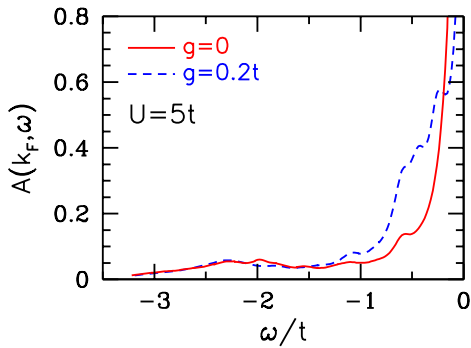}
\caption{(color online) DDMRG results on $A(k_F,\omega)$ of the
Hubbard-Holstein model for $\omega_0=0.2t$ and electron-phonon
couplings $g=0$ and $g=0.2t$. These results were obtained for
$U=5t$ and broadening $\gamma=0.1t$ on a 20-site chain with 12
electrons, for which $k_F\approx 6\pi/21$. This figure shows that,
at $T=0$, the electron-phonon interaction with $\omega_0=0.2t$
influences $A(k_F,\omega)$ for $|\omega|\lesssim 1.2t$. }
\label{fig4}
\end{figure}
In order to study the effects of the phonons within the presence
of the Coulomb interaction, we next present DDMRG results on
$A(k_F,\omega)$ for the Hubbard-Holstein model. In TTF-TCNQ,
inelastic neutron scattering experiments \cite{Pouget} have
revealed longitudinal acoustic and optical phonon modes. The
optical branch is weakly dispersive and has a frequency of about
$10meV$. Here, we present results for $U=5t$ and an Einstein
phonon mode with $\omega_0=0.2t$ and $g=0.2t$. In Fig. 4, we show
$A(k_F\approx 6\pi/21,\omega)$ for these parameters on a 20-site
chain with 12 electrons. For comparison, we also show results for
$g=0$. For the 20-site chain, the finite-size gap $\Delta_{FS}$
near $k_F$ is $\approx 0.3t$, and in this figure we have shifted
the spectrum by $\Delta_{FS}/2$ so that the main peak in
$A(k_F,\omega)$ occurs at $\omega=0$. Because of the finite
broadening $\gamma=0.1t$, we do not observe the opening of the CDW
gap. However, we see that the spectral weight at $\omega\approx 0$
is transferred to $-1.2t \ltsim \omega \ltsim -0.2t$. We have
performed similar calculations using $\omega_0=0.5t$, where
multi-phonon peaks are induced at energies $\approx -\omega_0$,
$-2\omega_0$, $-3\omega_0$, etc. with respect to the location of
the main peak. For physical values of $\omega_0\approx 10meV$, we
expect the changes in $A(k_F,\omega)$ at $T=0$ to occur at
$|\omega| \ll 1.3eV$. These results suggest that the
electron-phonon coupling is not particularly important for
investigating the spectral weight located at $-1.3eV \lesssim
\omega \lesssim -0.4eV$. However, it is still necessary to study
$A(k,\omega)$ of the 1D Hubbard-Holstein model at finite $T$,
since the electron-phonon interaction can influence the $T$
dependence of the magnetic correlations \cite{Scott,Torrance} and,
hence, the $T$ dependence of $A(k,\omega)$.

In summary, we have studied $A(k,\omega)$ of the 1D Hubbard model
in order to investigate the unusual $T$ dependence of the
photoemission intensity of TTF-TCNQ. We have also presented $T=0$
results for the 1D extended Hubbard and the 1D Hubbard-Holstein
models. We find that in the 1D Hubbard model the transfer of the
single-particle spectral weight takes place over a wide energy
range above a characteristic temperature, which is too high to
explain the ARPES data. We have shown that the long-range part of
the Coulomb repulsion can lead to a reduced value for $J_{eff}$
and, hence, to a better agreement with the ARPES data. Our results
on the Hubbard-Holstein model show that, at $T=0$, the
electron-phonon interaction does not influence $A(k_F,\omega)$
over the wide energy range observed by the ARPES. In conclusion,
these calculations give theoretical support to the notion that the
anomalous $T$ dependence of the photoemission spectrum of TTF-TCNQ
is due to the strong-correlation effects as suggested by Claessen
{\it et al.} \cite{Claessen}. However, we also emphasize that the
1D Hubbard model is not sufficient for explaining the unusual $T$
dependence, and at least the long-range part of the Coulomb
repulsion needs to be included.

We thank R. Claessen and F. Assaad for useful discussions, and R.
Claessen and M. Sing for permission to reproduce their ARPES data.
We also thank A. Abendschein and F. Assaad for sending us a copy
of their paper (Ref. \cite{Abendschein}) during the final stage of
this project. This work was supported by the NAREGI Nanoscience
Project and a Grant-in Aid for Scientific Research from the
Ministry of Education, Culture, Sports, Science and Technology of
Japan, and NEDO. One of us (N.B.) gratefully acknowledges support
from the Japan Society for the Promotion of Science and the
Turkish Academy of Sciences (EA-TUBA-GEBIP/2001-1-1).

\end{document}